\RequirePackage{lineno}
\documentclass[aps,prd, showpacs, twocolumn,superscriptaddress,groupedaddress]{revtex4}  
\usepackage{graphicx}  
\usepackage{dcolumn}   
\usepackage{bm}        
\usepackage{amssymb}   
\usepackage{amsmath}
\usepackage{subfigure}
\usepackage{array}
\newcommand{\PreserveBackslash}[1]{\let\temp=\\#1\let\\=\temp}
\newcolumntype{C}[1]{>{\PreserveBackslash\centering}p{#1}}
\newcolumntype{R}[1]{>{\PreserveBackslash\raggedleft}p{#1}}
\newcolumntype{L}[1]{>{\PreserveBackslash\raggedright}p{#1}}

\PassOptionsToPackage{pagewise}{lineno} 
\usepackage{lineno} 
\modulolinenumbers[1] 
   \newcommand*\patchAmsMathEnvironmentForLineno[1]{%
 \expandafter\let\csname old#1\expandafter\endcsname\csname #1\endcsname
 \expandafter\let\csname oldend#1\expandafter\endcsname\csname end#1\endcsname
 \renewenvironment{#1}%
 {\linenomath\csname old#1\endcsname}%
   {\csname oldend#1\endcsname\endlinenomath}}%
   \newcommand*\patchBothAmsMathEnvironmentsForLineno[1]{%
 \patchAmsMathEnvironmentForLineno{#1}%
   \patchAmsMathEnvironmentForLineno{#1*}}%
   \AtBeginDocument{%
 \patchBothAmsMathEnvironmentsForLineno{equation}%
 \patchBothAmsMathEnvironmentsForLineno{align}%
 \patchBothAmsMathEnvironmentsForLineno{flalign}%
 \patchBothAmsMathEnvironmentsForLineno{alignat}%
 \patchBothAmsMathEnvironmentsForLineno{gather}%
 \patchBothAmsMathEnvironmentsForLineno{multline}%
 }

\begin{document}
\title{Search for the rare decay of \bm{$\psi(3686)\rightarrow \Lambda_c^+ \overline{p} e^+ e^- + c.c.$} at BESIII}
\author{
  \small
M.~Ablikim$^{1}$, M.~N.~Achasov$^{9,d}$, S. ~Ahmed$^{14}$, M.~Albrecht$^{4}$, M.~Alekseev$^{55A,55C}$, A.~Amoroso$^{55A,55C}$, F.~F.~An$^{1}$, Q.~An$^{52,42}$, J.~Z.~Bai$^{1}$, Y.~Bai$^{41}$
% ...
\\
\vspace{0.2cm}
(BESIII Collaboration)\\
\vspace{0.2cm} {\it
$^{1}$ Institute of High Energy Physics, Beijing 100049, People's Republic of China\\
% ...
$^{4}$ Bochum Ruhr-University, D-44780 Bochum, Germany\\
% ...
$^{9}$ G.I. Budker Institute of Nuclear Physics SB RAS (BINP), Novosibirsk 630090, Russia\\
% ...
$^{14}$ Helmholtz Institute Mainz, Johann-Joachim-Becher-Weg 45, D-55099 Mainz, Germany\\
% ...
$^{41}$ Southeast University, Nanjing 211100, People's Republic of China\\
$^{42}$ State Key Laboratory of Particle Detection and Electronics, Beijing 100049, Hefei 230026, People's Republic of China\\
% ...
$^{52}$ University of Science and Technology of China, Hefei 230026, People's Republic of China\\
% ...
$^{55}$ (A)University of Turin, I-10125, Turin, Italy; (B)University of Eastern Piedmont, I-15121, Alessandria, Italy; (C)INFN, I-10125, Turin, Italy\\
% ...
$^{d}$ Also at the Novosibirsk State University, Novosibirsk, 630090, Russia\\
% ...
}
}

\date{\today}

\begin{abstract}
Based on a data sample of $(448.1\pm2.9)\times10^6~\psi(3686)$ decays collected with the BESIII experiment, a search for the flavor changing neutral current transition $\psi(3686)\rightarrow \Lambda_c^+ \overline{p} e^+ e^- + c.c.$ is performed for the first time. No signal candidates are observed and the upper limit on the branching fraction of $\psi(3686)\rightarrow \Lambda_c^+ \overline{p} e^+ e^-$ is determined to be $1.7\times10^{-6}$ at the 90\% confidence level. The result is consistent with expectations from the Standard Model,
and no evidence for new physics is found.
\end{abstract}

\maketitle

\section{Introduction}
Flavor changing neutral current (FCNC) transitions of heavy quarkonium are of great interest since they can provide indications for physics beyond the Standard Model (SM). In the framework of the SM, FCNC transitions are strongly suppressed by the Glashow, Iliopoulos and Maiani (GIM) mechanism \cite{GIM}.  The charm changing neutral current  (CCNC) decay of a charmonium state via a charm quark transition is only possible at the loop level. Furthermore, long-distance hadronic effects can contribute at the same level as the short-distance loop processes~\cite{Wang95}. The SM predictions of branching fractions (BFs) for FCNC decays range from $10^{-10}$ to $10^{-14}$~\cite{SL94, Wang08}.   However, some new physics models such as the Topcolor model~\cite{Hill95},  the minimal supersymmetric SM with R-parity violation~\cite{Aulakh95} and the two Higgs doublet model~\cite{Glashow77} predict the BFs of the same FCNC decays to be two to three orders of magnitude larger. 
Any observation of a FCNC decay of charmonium states with the current experimental sensitivity would be would be clear evidence for physics beyond the SM~\cite{Zhang01, Datta99}. 

 \begin{figure}
\centering 
\includegraphics[width=0.9\columnwidth]{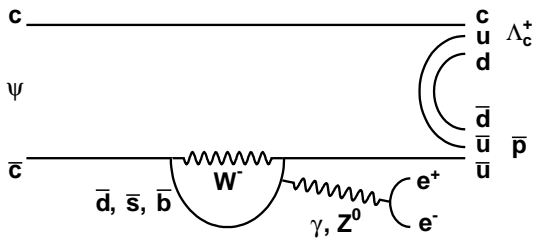}
\caption{Feynman diagram for the CCNC transition of $\psi(3686)\rightarrow \Lambda_c^+ \overline{p} e^+ e^-$.}
\label{fig:feynman} 
\end{figure} 

The Feynman diagram of the decay $\psi(3686)\rightarrow \Lambda_c^+ \overline{p} e^+ e^-$ at loop level is shown in Fig.~\ref{fig:feynman}.  In this paper we present a search for the rare decay of $\psi(3686)\rightarrow \Lambda_c^+ \overline{p} e^+ e^-$ using a sample of $(448.1\pm2.9)\times10^6 ~\psi(3686)$ events~\cite{Zhiyong17} collected by the BESIII detector.  Charged conjugation is implied throughout the paper.

\section{BESIII DETECTOR and Monte Carlo Simulation}
\label{sec:BESIII}
The Beijing Electron Positron Collider II (BEPCII) is a symmetric $e^+ e^-$ collider located at the Institute of High Energy Physics (IHEP) in Beijing. The accessible center-of-mass energy ($\sqrt{s}$) ranges from 2.0 to 4.6~GeV. At $\sqrt{s}$ = 3.773~GeV, a maximum luminosity of  $1.0\times10^{33}~\text{cm}^{-2}\text{s}^{-1}$ is achieved.  The BESIII detector has a geometrical acceptance of 93\% of the solid angle.  The main drift chamber (MDC) provides momentum measurements of charged tracks with a precision of 0.5\% at 1~GeV$/c$ and measurements of the energy loss ($\text{d}E/\text{d}x$) with a precision of 6\%.  The time-of-flight (TOF) system consists of plastic scintillators and provides a measurement of the flight time with a resolution of 80 and 110~ps for the barrel and end-cap parts of the detector, respectively.  The combined information from $\text{d}E/\text{d}x$ and TOF is used to identify particle species of charged tracks. The electromagnetic calorimeter (EMC) is used to measure the energy of photons with a resolution of 2.5\% and 5.0\% at 1~GeV for the barrel and end-cap  parts, respectively. The muon counter (MUC) system consists of resistive plate chambers and measures the position of muon tracks with a precision better than 2~cm. Further information on the detector can be found in Ref.~\cite{bes310}. 

Monte Carlo (MC) simulation is used to optimize selection criteria, determine the reconstruction efficiency and estimate the possible backgrounds. The $e^+e^-$ collision and the production of the charmonium resonance are simulated using {\sc kkmc}~\cite{KKMC} and the subsequent particle decays using {\sc evtgen}~\cite{Lange01} for the known decay modes. The remaining unknown decay modes are simulated using the {\sc lundcharm} model~\cite{chen00}. The simulation of the particle interactions with the detector is based on {\sc geant4}~\cite{Geant4}. An `inclusive' MC sample of $506\times 10^6$ generic $\psi(3686)$ decays is used to study possible backgrounds.  
An exclusive signal MC sample $\psi(3686)\rightarrow \Lambda_c^+ \overline{p} e^+ e^-$ is generated to determine the reconstruction efficiency. The signal MC sample is generated using a vector meson dominance (VMD) model~\cite{VMD0, VMD1, VMD2}, where the $e^+ e^-$ pair in the final state is produced from a virtual photon decay.  
The VMD model is also implemented in Refs.~\cite{etap, D0}.  Due to the lack of data, the corresponding form factor of $\psi(3686)\rightarrow \Lambda_c^+ \overline{p} e^+ e^-$ in the VMD model is taken from the decay $\rho \rightarrow \pi^+ \pi^- e^+ e^-$~\cite{VMDrho}, where the form factor with four-momentum transfer squared ($Q^2$) dependence is denoted by the hidden gauge model as described in Ref.~\cite{VMD2}. In the VMD model, the width of vector meson is introduced to  eliminate the singularities of the mass of the vector meson. The decay  $\Lambda_c^+ \rightarrow p K^- \pi^+$ is simulated using the model described in Ref.~\cite{Lambdacp}, in which interference between the non-resonant and resonant contributions is included.

\section{Event Selection}

\subsection{Charged track selection}
The decay $\psi(3686)\rightarrow \Lambda_c^+ \overline{p} e^+ e^-$ with $\Lambda_c^+ \rightarrow p K^- \pi^+$ is reconstructed with six charged tracks with zero net charge.  Each charged track is required to be within the acceptance of the MDC (polar angle $|\cos\theta|<0.93$). Furthermore, we require that the point of closest approach is separated from the interaction point by less than 10~cm along and 1~cm perpendicular to the beam direction.   For each track candidate, confidence levels for different particle hypotheses (proton, kaon, pion and electron) are calculated using $\text{d}E/\text{d}x$  and  TOF information. The charged tracks are assigned the particle type corresponding to the highest confidence level. No additional charged tracks are allowed besides the six candidate tracks.

\subsection{Kinematic fit}
A vertex fit is applied to the selected track candidates and is required to converge. The four momenta of the tracks are updated according to the fitted values. Furthermore, a four-constraint (4C) kinematic fit imposing energy-momentum conservation under the hypothesis of $\psi(3686)\rightarrow p \overline{p} K^- \pi^+  e^+ e^-$ is applied to improve the mass resolution and suppress background. The $\chi^2$ of the 4C kinematic fit is required to be less than 200. 

\subsection{Further background suppression}
The possible background contamination from other $\psi(3686)$ decays is studied with the inclusive MC sample. There are only 29 simulated events that survive the above selection criteria.  These are dominated by the processes $\psi(3686) \rightarrow \gamma \chi_{cJ}$, $\chi_{cJ} \rightarrow p K^- \overline{\Lambda}$ and $\psi(3686) \rightarrow \overline{\Lambda} K^{*-} p$, $K^{*-}\rightarrow K^- \pi^0$, where the selected $e^+e^-$ pair is from $\gamma$ conversion (through interactions with the detector material) or from $\pi^0$ Dalitz decays. The above background processes contain the intermediate state $\overline{\Lambda}$, and are rejected by requiring the invariant mass of $\overline{p} \pi^+$ ($M_{\overline{p} \pi^+}$)  to be greater than 1.13~GeV$/c^2$.

The possible backgrounds from the continuum QED and two-photon processes are examined using a data sample of 2.93 $\rm fb^{-1}$ collected at $\sqrt{s}=3.773$~GeV~\cite{psi3770}. No events with the invariant mass  of $pK^-\pi^+$ ($M_{pK^-\pi^+}$) ranging between 2.0 and 2.4~GeV$/c^2$ survive.  It is therefore concluded that the backgrounds from the QED and two-photon processeses are negligible. 

\section{Systematic uncertainty}
In the measurement of the BF of the decay $\psi(3686)\rightarrow \Lambda_c^+ \overline{p} e^+ e^-$, systematic uncertainties arise from the following sources:  

\begin{itemize} 
\item [(I)] The \textbf{total number of $\bm{\psi(3686)}$ events} is determined by a measurement of inclusive hadronic final states~\cite{psi_number} with an uncertainty of 0.6\%.

\item [(II)]  The difference between data and MC simulation in efficiencies of \textbf{track reconstruction and particle identification (PID)} are estimated using the control samples of $\psi(3686) \rightarrow \pi^+ \pi^- J/\psi$ with $J/\psi \rightarrow e^+ e^-$ and $J/\psi \rightarrow p K^- \Lambda + c.c.$.  The systematic uncertainties are estimated to be less than 1.0\% per track for track reconstruction and PID, individually~\cite{systPID}. Due to the low momentum of leptons, we further use the radiative Bhabha scattering events ($e^+e^- \rightarrow \gamma e^+e^-$) to study the systematic uncertainties for the leptons. The lepton tracks with momentum lower than 300 MeV$/c$ are selected as the control sample. The difference in efficiencies between the data and MC sample generated at  $\sqrt{s} =$ 3.097 GeV is assigned as the systematic uncertainty. The systematic uncertainties of efficiency for the lepton tracking and PID are estimated to be less than 2.5\%, individually. 

\item [(III)]  The difference between data and MC simulation due to the \textbf{4C kinematic fit} is estimated using the control sample of $\psi(3686) \rightarrow \pi^+ \pi^- J/\psi, J/\psi \rightarrow p \overline{p} \pi^+ \pi^-$. An agreement better than 1.0\% is found and we assign 1.0\% as the systematic uncertainty. 

\item [(IV)]  The \textbf{BF of $\bm{\Lambda_c^+ \rightarrow p K^- \pi^+}$} is an external input parameter and quoted from Ref.~\cite{BRlamb} to be $(6.35\pm0.33)$\%. The relative uncertainty of 5.2\% is taken as the systematic uncertainty. 

\item [(V)]  The signal is examined in the $M_{pK^-\pi^+}$  distribution ranging from 2.25 to 2.32~GeV$/c^2$. An alternative \textbf{signal region} ranging from 2.27 to 2.30 GeV$/c^2$ is also used to examine the signal and the corresponding change of signal efficiency, 4.0\%, is assigned as the systematic uncertainty.

\item [(VI)]  The systematic uncertainty due to \textbf{the requirement on the \bm{ $M_{\overline{p}\pi^+}$} distribution} is studied using a control sample of $e^+ e^- \rightarrow \Lambda_c^+ \overline{\Lambda}^-_c$ with $\Lambda_c^+$ decaying into $p K^- \pi^+$ at $\sqrt{s}~=$ 4.6~GeV with an integrated luminosity of $567~\rm pb^{-1}$~\cite{c4600}. By applying the same $M_{\overline{p}\pi^+}$ selection requirement, we calculate the corresponding efficiency as the ratio of the events with and without the selection requirement. The efficiency difference between data and MC simulation, 1.0\%, is assigned as the systematic uncertainty. 

\item [(VII)] We study the influence of the \textbf{physics model} of the decay $\psi(3686)\rightarrow \Lambda_c^+ \overline{p} e^+ e^-$ by changing the decay model to an extreme model and a phase space model. In the extreme model, we assume an additional intermediate decay of $\psi(3686)\rightarrow X \overline{p}$, where the polar angle distribution of $\overline{p}$ follows $1+\cos^2\theta$ and $X$ decays to $ \Lambda_c^+ e^+ e^-$ according to a VMD model. The difference in the signal detection efficiency is 34.3\% which is mainly due to  the different geometrical acceptance for the events and the difficulty in finding low momentum leptons with respect to the nominal physics model.   In the phase space model, we assume a uniform phase space distribution for signal, and the resulting difference in efficiency with respect to the nominal value is found to be 8.3\%. We assign 34.3\% as the systematic uncertainty.
\end{itemize} 

A summary of all systematic uncertainties is given in Table~\ref{tab:syst}. The total uncertainty is 37.2\%, which is the quadrature sum of the individual values.

 \begin{table}[thbp]
  \caption[systematic uncertainty results]{Overview of systematic uncertainties.}
  \label{tab:syst}
  \begin{tabular}{lcccccc} \hline
   \hline       Sources  & Systematic uncertainty  (\%)\\ 
    \hline      Number of $\psi(3686)$ decays   & 0.6 \\
                   Track reconstruction  &  9.0 \\
                    Particle identification & 9.0 \\ 
                    4C kinematic fit & 1.0 \\
                    BF of $\Lambda_c^+ \rightarrow p K^- \pi^+$ & 5.2 \\
                    Signal region     & 4.0 \\
                    $M_{p \pi^-}/M_{\overline{p}\pi^+}$ criteria  & 1.0 \\
                    Physics model  & 34.3  \\\hline
                    Total  & 37.2 \\
                   \hline
    \hline
  \end{tabular}
\end{table}
 
\section{Result} 
\begin{figure}[!t]
   \includegraphics[width=0.9\columnwidth]{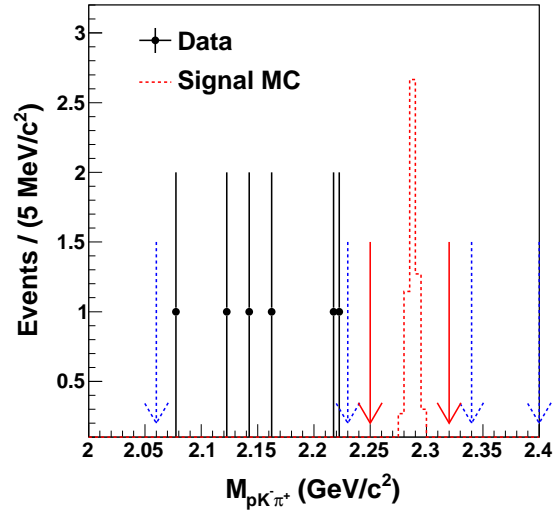}
     \renewcommand{\figurename}{Fig.}
   \caption{Distribution of $M_{pK^-\pi^+}$ for the data (dots with
     error bars) and signal MC sample (dashed histogram).  The signal
     MC is scaled arbitrarily.  The regions between the left (right) two blue dashed and middle two red solid arrows represent the sideband and signal regions, respectively.}
   \label{fig:mpkpidata}
 \end{figure}

The number of signal events is determined by examining the $\Lambda^+_c$ signal in the $M_{pK^-\pi^+}$ distribution, which is shown in Fig.~\ref{fig:mpkpidata}.  No events survive within the signal region ranging from 2.25 to 2.32 GeV/$c^2$.  The potential background in the signal region is estimated using events in the $M_{pK^-\pi^+}$ sideband regions, which are defined as [2.06, 2.23]~GeV$/c^2$ and [2.34, 2.40]~GeV$/c^2$. The estimated number of background events is 1.5, assuming a uniform distribution of background in the $M_{pK^-\pi^+}$ distribution. We also estimate the number of background events to be zero using the inclusive MC sample and the data sample with $\sqrt{s}=3.773$~GeV. As no candidate events are found in the signal region, the estimated number of background events is determined to be $0\pm1.5$ events.  Using the Rolke method~\cite{Rolke05, ROOT17}, an upper limit $N_{\rm up}$ of 47.3 produced events at the 90\% confidence level (C.L.) is obtained.  This upper limit takes into account the number of background events, the systematic uncertainty, and the detection efficiency~(7.21\%). The number of signal events is assumed to follow a Poisson distribution, and the signal detection efficiency and the number of background events are assumed to follow Gaussian distributions with widths given by the corresponding uncertainties.  The upper limit on the BF ($\mathcal{B}$) of the decay $\psi(3686)\rightarrow \Lambda_c^+ \overline{p} e^+ e^- + c.c.$ is calculated to be $1.7\times 10^{-6}$ using the following formula: 
\begin{equation}
\label{eq:BRcal}
\mathcal{B} \le \frac{N_{\rm up}}{N_{\psi(3686)} \times \text{BF}(\Lambda_c^+ \rightarrow p K^- \pi^+) }, 
\end{equation}
where  $N_{\psi(3686)}$ is the number of $\psi(3686)$ decays and $\text{BF}(\Lambda_c^+ \rightarrow p K^- \pi^+)$ is the BF of the decay $\Lambda_c^+ \rightarrow p K^- \pi^+$~\cite{BRlamb}.

\section{SUMMARY}
\label{sec:summary}
The search for the FCNC decay $\psi(3686)\rightarrow \Lambda_c^+ \overline{p} e^+ e^- + c.c.$  is performed for the first time using a sample of $(448.1\pm2.9)\times10^6$  $\psi(3686)$ decays.  No signal events are observed and the upper limit on the BF at the 90\% C.L. is determined to be $1.7\times10^{-6}$.  The result is within the expectations of the SM, and no evidence for new physics is found.

\section{ACKNOWLEDGMENT}
The BESIII collaboration thanks the staff of BEPCII and the IHEP computing center for their strong support. This work is supported in part by National Key Basic Research Program of China under Contract No. 2015CB856700; National Natural Science Foundation of China (NSFC) under Contracts Nos. 11375204, 11235011, 11335008, 11425524, 11625523, 11635010; the Chinese Academy of Sciences (CAS) Large-Scale Scientific Facility Program; the CAS Center for Excellence in Particle Physics (CCEPP); the Collaborative Innovation Center for Particles and Interactions (CICPI); Joint Large-Scale Scientific Facility Funds of the NSFC and CAS under Contracts Nos. U1332201, U1532257, U1532258; CAS Key Research Program of Frontier Sciences under Contracts Nos. QYZDJ-SSW-SLH003, QYZDJ-SSW-SLH040; 100 Talents Program of CAS; National 1000 Talents Program of China; INPAC and Shanghai Key Laboratory for Particle Physics and Cosmology; German Research Foundation DFG under Contracts Nos. Collaborative Research Center CRC 1044, FOR 2359; Istituto Nazionale di Fisica Nucleare, Italy; Koninklijke Nederlandse Akademie van Wetenschappen (KNAW) under Contract No. 530-4CDP03; Ministry of Development of Turkey under Contract No. DPT2006K-120470; National Natural Science Foundation of China (NSFC) under Contracts Nos. 11505034, 11575077; National Science and Technology fund; The Swedish Research Council; U. S. Department of Energy under Contracts Nos. DE-FG02-05ER41374, DE-SC-0010118, DE-SC-0010504, DE-SC-0012069; University of Groningen (RuG) and the Helmholtzzentrum fuer Schwerionenforschung GmbH (GSI), Darmstadt; WCU Program of National Research Foundation of Korea under Contract No. R32-2008-000-10155-0.


\begin{thebibliography}{}
\bibitem{GIM}
S.~L.~Glashow {\it et al.}, Phys. Rev. D {\bf 2}, 1285 (1970).
\bibitem{Wang95}
Y.~M.~Wang {\it et al.}, J. Phys. G {\bf 36}, 105002  (2009).  
\bibitem{SL94}
M.~A.~Sanchis-Lonzano, Z. Phys. C {\bf 62}, 271 (1994).
\bibitem{Wang08}
Y.~M.~Wang {\it et al.}, Eur. Phys. J. C {\bf 54}, 107 (2008).
\bibitem{Hill95}
C.~Hill, Phys. Lett. B {\bf 345}, 483 (1995).
\bibitem{Aulakh95}
C.~S.~Aulakh and R.~N.~Mohapatra, Phys. Lett. B {\bf 119}, 136 (1982).
\bibitem{Glashow77}
S.~Glashow and S.~Weinberg, Phys. Rev. D {\bf 15}, 1858 (1977).
\bibitem{Zhang01}
X.~Zhang, arXiv: hep-ph/0010105 (2000).
\bibitem{Datta99}
A.~Datta {\it et al.}, Phys. Rev. D {\bf 60}, 014011 (1999).
\bibitem{Zhiyong17}
M.~Ablikim \textit{et al.} (BESIII Collaboration), arXiv:1709.03653, submitted to Chin. Phys. C.
\bibitem{bes310}
M.~Ablikim {\it et al.} (BESIII Collaboration), Nucl. Instrum. Meth. A {\bf 614}, 345 (2010).
\bibitem{KKMC}
S.~Jadach {\it et al.}, Comput. Phys. Commun. {\bf 130}, 260 (2000); Phys. Rev. D {\bf 63}, 113009 (2001).
\bibitem{Lange01}
D.~J.~Lange, Nucl. Instrum. Meth. A {\bf 462}, 152 (2001); R.~G.~Ping, Chin. Phys. C {\bf 32}, 599 (2008).
\bibitem{chen00}
J.~C.~Chen, G.~Huang, X.~Qi, D.~Zhang, and Y.~Zhu, Phys. Rev. D 62, 034003 (2000).
\bibitem{Geant4}
S.~Agostinelli {\it et al.} (GEANT4 Collaboration), Nucl. Instrum. Meth. A {\bf 506}, 250 (2003).
\bibitem{VMD0}
J. ~J. ~Sakurai, Phys. Rev. Lett. {\bf 22}, 981 (1969).
\bibitem{VMD1}
 V.~M.~Budnev and V.~A.~Karnakov, Pisma Zh. Eksp.Teor. Fiz. {\bf 29}, 439 (1979).
 \bibitem{VMD2}
 Z.~Y.~Zhang, L.~Q.~Qin and S.~S.~Fang, Chin.Phys. C {\bf 36},  926 (2012).  
\bibitem{etap}
M.~Ablikim {\it et al.} (BESIII Collaboration), Phys. Rev. D {\bf 87}, 092011 (2013). 
\bibitem{D0}
M.~Ablikim {\it et al.} (BESIII Collaboration),  Phys. Rev. D  {\bf 96}, 111101 (2017). 
\bibitem{VMDrho}
T.~Petri, arXiv:nucl-th/1010.2378 (2010).
\bibitem{Lambdacp}
M.~Ablikim {\it et al.} (BESIII Collaboration), Phys. Rev. Lett. {\bf 116}, 052001 (2016). 
\bibitem{psi3770}
M.~Ablikim {\it et al.} (BESIII Collaboration), Phys. Lett. B {\bf 753}, 629 (2016).
\bibitem{psi_number}
 M.~Ablikim {\it et al.} (BESIII Collaboration),  arXiv: 1709.03653 (2017). 
\bibitem{systPID}
M.~Ablikim {\it et al.} (BESIII Collaboration), Phys. Rev. D {\bf 86} 032008 (2012);  Phys. Rev. D {\bf 87} 112007 (2013).
\bibitem{BRlamb}
C.~Patrignani {\it et al.} (Particle Data Group), Chin. Phys. C {\bf 40}, 100001 (2016).
\bibitem{c4600}
 M.~Ablikim {\it et al.} (BESIII Collaboration),  Chin. Phys. C {\bf 39}, 093001 (2015).
\bibitem{Rolke05}
W.~A.~Rolke, A.~M.~Lopez, J.~Conrad, Nucl. Instrum. Meth. A {\bf 551}, 493 (2005).
\bibitem{ROOT17}
R.~Brun and F.~Rademakers, Nucl. Instrum. Meth.  A {\bf 389}, 81 (1997).
\end{thebibliography}
\end{document}